\documentclass[apjl]{emulateapj}
\usepackage{epsfig}
\begin{document}

\title{Radial disk heating by more than one spiral density wave}

\author{I.~Minchev and A.~C.~Quillen}
\affil{(Department of Physics and Astronomy,
University of Rochester)
}

\begin{abstract}

We consider a differentially rotating, 2D stellar disk perturbed by two steady 
state spiral density waves moving at different patterns speeds.
Our investigation is based on direct numerical 
integration of initially circular test-particle orbits. 
We examine a range of spiral strengths and spiral speeds and show that 
stars in this time dependent gravitational field 
can be heated (their random motions increased). This is particularly 
noticeable in the simultaneous propagation of a 2-armed
spiral density wave near the corotation resonance (CR), and a weak 
4-armed one near the inner and outer 4:1 Lindblad resonances.
In simulations with 2 spiral waves moving at
different pattern speeds we find:
(1) the variance of the radial velocity, ${\sigma_R}^2$, exceeds the 
sum of the variances measured from simulations with each 
individual pattern; 
(2) $\sigma_R^2$ can grow with time throughout the entire simulation;
(3) $\sigma_R^2$ is increased over a wider range of radii compared to that seen
with one spiral pattern;
(4) particles diffuse radially in real space whereas they don't when
only one spiral density wave is present.
Near the CR with the stronger, 2-armed pattern, test particles are observed 
to migrate radially. These effects take place at or near resonances of both 
spirals so we interpret them as the result of stochastic motions. 
This provides a possible new mechanism for increasing 
the stellar velocity dispersion in galactic disks.
If multiple spiral patterns are present in the Galaxy we 
predict that there should be large variations in the stellar velocity
dispersion as a function of radius.

\end{abstract}


\section{Introduction}

The observed correlation between the ages and velocity dispersions of solar
neighborhood stars \citep{eggen,dennis,wielen}
has been a subject of study since the work of 
\citet{ss_51,ss_53}.
They established that scattering of
stars from initially circular orbits into more eccentric
and inclined orbits was a likely explanation for the increase in
velocity dispersion, $\sigma$, with age $t$.
They suggested that massive gas clouds (then undetected)
were responsible.
Molecular clouds were thought to be the sole scattering agents
(e.g., \citealt{mihalas}) until \citet{lacey84}
showed that the observed ratio of the dispersion
in the direction perpendicular to the Galactic plane
and that toward the Galactic center, ${\sigma_z / \sigma_R}$,
was too low to be consistent with the predictions from this scattering
process. There was also a discrepancy between the predicted and
then measured relation between $\sigma$ and $t$: if $\sigma \propto t^\alpha$,
models predicted $\alpha \sim 0.20$, while observations of the time
yielded $\alpha \sim 0.5 $ \citep{wielen}.
This resulted in the development of models that incorporated the
heating of the stellar disk from transient spiral structure
\citep{barbanis,sell_carl,carl_sell} in
addition to scattering from molecular clouds \citep{jenkins,jenkins92}.
Other proposed models for the heating of stars include scattering by halo black 
holes \citep{lacey_ost} or dark clusters \citep{carr},
giant molecular clouds and halo black holes \citep{han02,han04}, and 
infall of satellite galaxies \citep{vel}.

We define a steady state spiral pattern to be a spiral perturbation moving at a 
constant angular velocity with a fixed amplitude and a fixed pitch angle.
An individual steady state spiral arm pattern is not expected to
increase the velocity dispersion of a stellar population 
(e.g., \citealt{B+T}), though this may be violated 
at resonances (e.g., \citealt{cont81, combes, fux}). Consequently,
models used to explain the age/dispersion relation rely on transient spiral 
structure.  
In the solar neighborhood, at a distance of 8 kpc from
the Galactic center and with an angular rotation speed of 220 km/s
one rotation period for a star corresponds to about 0.25 Gyrs.
Thus, only about 40 rotation periods have taken place during 
the lifetime of the Milky Way disk.  Because of the small number of periods,
it is difficult to apply a diffusive theory to the problem of disk heating.

\citet{quillen_chaos} pointed out that when perturbations at two pattern speeds,
such as a bar and a spiral density wave are present in the disk,
the stellar dynamics can be stochastic, particularly near 
resonances associated with one of the patterns.
While this paper focused on spiral structure near the Galactic bar's outer
Lindblad resonance, here we consider the more general possibility that
there could be more than one wave present in the disk.
There is also observational evidence for this in other galaxies, 
such as asymmetries in the spiral structure
(e.g., \citealt{henry}). 
By expanding galaxy images in Fourier components,
\citet{elmegreen92} noted that many galaxies exhibit
hidden three armed components and suggested that multiple
spiral density waves can propagate simultaneously in galaxy disks.
\citet{rix93}, in their Fourier expansion of the near-infrared images
of M51 saw both $m=1$ and $m=3$ components as well as the
dominant 2-armed structure.
In their recent work \citet{naoz05}
found evidence for multiple spiral patterns in our home galaxy. By studying the nearby 
spiral arms they find that the Sagittarius-Carina arm is a superposition of 
two features, moving at different pattern speeds.
Clumps containing old stars in the Solar neighborhood velocity distribution
\citep{dehnen98},  
may be interpreted in terms of multiple stellar density waves traveling
in our vicinity in the Milky Way disk \citep{desimone}. 
Theoretically this situation corresponds to scenarios
that include multiple and transient spiral density waves
(e.g., \citealt{toomre,fuchs}) rather than
those that focus on a single dominant quasi-steady mode 
(e.g., \citealt{lowe}).

In this paper we are interested in the effect on the radial velocity dispersion
and on the orbits of stars caused by the simultaneous propagation of two steady 
state spiral patterns in a combination of (strong) 2-armed + (weak) 4-armed, 
where the second pattern moves at an angular speed different than the first one.
This situation has been motivated by \citet{lepine}, \citet{drimmel}, 
\citet{naoz05}, and \citet{q+m} where a combination of 2-armed and a 4-armed 
spiral waves has been considered in the Solar Neighborhood.
 
We study the kinematic and spatial 
evolution of stars in a 2D disk by following the trajectories of 
non-self-gravitating particles under the gravitational potential of the 
background (disk and halo) and the 2 spiral density waves just described. 
We explore a range of pattern speed combinations and look for an increase 
in the radial velocity dispersion of particles.

Our simulation model is described in \S \ref{sec:model}. We describe our 
simulations in \S \ref{sec:onepattern}-\S \ref{sec:orbits}. A summary and
discussion follows in \S \ref{sec:sum}.  

\section{Numerical model}
\label{sec:model}

\subsubsection{Notation and units used}

For simplicity we work in units in which the galactocentric distance 
to the annulus in which stars are distributed is $R_0=1$, 
the circular velocity at $R_0$ is $V_0=1$, and the angular velocity
of stars is $\Omega_0 = V_0/R_0=1$. Actually, since we assume a flat rotation 
curve throughout the paper the initial circular velocity is $V_0=1$ everywhere.
One orbital period is $2\pi R_0 / \Omega_0 = 2\pi$. In our code time is in units 
of $1/\Omega$ whereas in the figures it is given in units of
orbital periods at $R_0$. The velocity vector of a star
is $(u,v)$, where $u,v$ are the radial and tangential velocities in a reference
frame rotating with $V_0$. Consequently, the tangential velocity of a star in
an inertial reference frame is $V_0 + v$.

In the definitions that follow the subscript $a$ refers to the primary, 2-armed
spiral wave; similarly the subscript $b$ refers to the secondary, 4-armed 
one. The azimuthal wavenumbers of each spiral wave, $m_a$, $m_b$, are integers 
corresponding to the number of arms.
The amplitudes of the spiral wave gravitational potential perturbations are 
denoted as $\epsilon_a$, $\epsilon_b$. 
The spiral pattern frequencies of each spiral wave are $\Omega_{s,a}$ and 
$\Omega_{s,b}$.
The parameters $\alpha_a$, $\alpha_b$ are 
related to the pitch angles of the spiral waves, $p_a$, $p_b$, as 
$\alpha = m \cot(p)$, negative for trailing spirals with rotation 
counterclockwise. In this paper we only consider trailing spiral waves.

We write the Hamiltonian of a star as $H=H_0+H_1$ where $H_0$ is the 
unperturbed, axisymmetric part and $H_1$ is a small perturbation due to 
the 2-armed, 4-armed, or both spiral waves. We write $H_1=\Phi_a+\Phi_b$
where $\Phi_a,\Phi_b$ are the gravitational potential perturbations due to 
each pattern (see next section).

In general we refer to the radial velocity dispersion as $\sigma_R$. However,
we also define 3 other symbols referring to the type of perturber we use in
the Hamiltonian. Considering the above described form of  
$H_1$, if 
$\Phi_a=0$ then $\sigma_R\longrightarrow\sigma_{R,b}$ (only 4-armed pattern 
present); 
if $\Phi_b=0$, $\sigma_R\longrightarrow\sigma_{R,a}$ (only 2-armed pattern 
present);
finally, if both $\Phi_a,\Phi_b\neq0$ then 
$\sigma_R\longrightarrow\sigma_{R,ab}$ (both spiral waves present).

\subsection{Equations of motion}

We consider the $2D$ motion of a test particle in the mid-plane of a galaxy in 
an inertial reference frame.
In plane polar coordinates, $R, \phi$, the Hamiltonian of a star can be written as 
\begin{equation}
\label{eq:ham}
H(R,\phi,p_R, p_{\phi},t)=H_0(R,p_R, p_{\phi}) + H_1(R, \phi, t)
\end{equation}
The first term on the right hand side is the unperturbed axisymmetric 
Hamiltonian
\begin{equation}
H_0(R,p_R, p_{\phi}) = {p_R^2 \over 2} + {p_{\phi}^2 \over {2 R^2}} + \Phi_0(R)  
\end{equation}
where we assume the axisymmetric background potential due to the disk and 
halo has the form $\Phi_0(R)=V_0\log(R)$, corresponding to a flat rotation 
curve. Energy and angular momentum for $H_0$ are conserved. 

In the case of one periodic perturbation there is still a conserved
quantity in the reference frame rotating with the pattern. That is the Jacobi 
integral, $J=E-L\Omega_s$, where $E$ is the energy of the particle, $L$ is its
angular momentum, and $\Omega_s$ is the pattern angular velocity.
However, upon adding a second spiral density perturbation $H$ does not admit
any isolating integrals.

The perturbation is
$H_1(R, \phi, t)=\Phi_a(R, \phi, t) + \Phi_b(R, \phi, t)$ where the 2 terms
on the right hand side are the gravitational potentials due to each spiral 
wave. We define the subscripts $a,b$ to denote the 2- and 4-armed spiral 
patterns, respectively. The perturbations are discussed in the next subsection. 

Hamilton's equations applied to Eq. \ref{eq:ham} 
give us the equations of motion

\begin{eqnarray}
\dot p_R = {p_{\phi}^2\over{R^3}}-\Phi_0^\prime-\Phi_a^\prime-\Phi_b^\prime  \\ 
\dot p_{\phi}=-{\partial \Phi_a\over\partial\phi} -{\partial\Phi_b\over\partial\phi}
\end{eqnarray}

In the first of these equations the primes denote partial derivatives with 
respect to $R$.

\subsubsection{Perturbation from spiral structure}

We treat the spiral patterns as small perturbations to the axisymmetric 
model of the galaxy 
by viewing them as quasi-steady density waves in accordance with the Lin-Shu 
hypothesis \citep{lin-shu}. Each spiral wave gravitational potential 
perturbation is expanded in Fourier components as
\begin{equation}
\label{eq:phi}
\Phi(R, \phi, t)=\sum_m \epsilon_m
\exp \left[i (\alpha \ln{R} - m(\phi-\Omega_s t - \phi_m ))\right]
\end{equation}
and its corresponding surface density is
\begin{equation}
\label{eq:sigma}
\Sigma(R, \phi, t)=\sum_{m} \Sigma_m
\exp \left[i (\alpha \ln{R} - m(\phi-\Omega_s t - \phi_m ))\right]
\end{equation}
We assume that the amplitudes, $\epsilon_m, \Sigma_m$, and the pitch angles 
are nearly constant with radius. 
The strongest term for a 2-armed spiral is the 
$m=2$ term and similarly $m=4$ for a 4-armed structure. Thus only the terms 
corresponding to those $m$ values are retained. In this paper we only consider 
2- and 4-armed spiral waves, thus in our notation $m_a=2$ and $m_b=4$. Upon taking 
the real parts of Eqs. \ref{eq:phi} and \ref{eq:sigma} 
the perturbation due to the 2-armed spiral density wave becomes

\begin{equation}
\Phi_a(R,\phi,t) = 
\epsilon_a \cos{(\alpha_a \ln{R}-m_a(\phi-\Omega_{s,a} t-\phi_{0,a}))}.
\end{equation}
Changing the subscript $a$ to $b$ in the above equation gives the 4-armed 
perturbation. The direction of rotation is with 
increasing $\phi$ and $\alpha_a, \alpha_b < 0$ ensure that each pattern is 
trailing. 

The recent study of \citet{vallee05} provides a good summary of
the many studies which have used observations to map the Milky Way disk.
Cepheid, HI, CO and far-infrared observations suggest that the Milky Way
disk contains a 4-armed tightly wound structure, whereas \citet{drimmel}
have shown that the near-infrared observations are consistent with a dominant
2-armed structure. The dominant 2-armed and
weaker 4-armed structure was previously proposed by \citet{amaral}.
We adopt the same configuration in our model - a primary 2-armed and a weaker 
4-armed spiral wave perturbations. What is different and new in our model is the 
assumption that there is a non-zero relative angular velocity 
between the 2 spiral patterns. This introduces an additional parameter - 
the pattern speed of the secondary spiral wave.

In general, an individual stellar orbit is affected by the entire galaxy. 
However, if tight winding of spiral arms is assumed only local gravitational 
forces need be considered. \citet{ma} find $kR$ for several spiral galaxies 
of various Hubble types to be $>$ 6, thus the tight-winding, or WKB 
approximation, is often appropriate. In the above expression $k$ is the 
wave-vector and $R$ is the radial distance from the Galactic Center (GC), 
related to the pitch angle $p$ through 
$\cot(p)=|kR/m|$. This gives $\alpha$ in terms of $k$ as $\alpha\approx|kR|$. 
In the WKB approximation the amplitude of the potential
perturbation Fourier component is related to the density perturbations in the 
following way
\begin{equation}
\epsilon_a \sim { - 2 \pi G  \Sigma_0 S_a R_0 \over |\alpha_a| V_0^2}
\end{equation}
(\citealt{B+T}). The above equation is in units of $V_0^2$.
$\Sigma_a$ is the amplitude of the primary, 2-armed-spiral-wave mass surface 
density and $S_a\approx \Sigma_a/\Sigma_0$. A similar relation holds for the 
second spiral pattern.

\subsubsection{Resonances}
 
Of particular interest to us are the values of $\Omega_s$ which place the 
spiral waves near resonances. The Corotation Resonances (CR) occurs when the 
angular rotation rate of stars equals that of the spiral pattern.
Lindblad Resonances (LRs) occur when the 
frequency at which a star feels the force due to a spiral arm coincides with 
the star's epicyclic frequency, $\kappa$. As one moves inward or outward from 
the corotation circle the relative frequency at which a star encounters a 
spiral arm increases. There are 2 values of $R$ for which this frequency is 
the same as the epicyclic frequency and this is where the Outer Lindblad 
Resonance (OLR) and the Inner Lindblad Resonance (ILR) are located.

Quantitatively, LRs occur when $\Omega_s=\Omega_0 \pm \kappa/m$. 
The negative sign corresponds to the 
ILR and the positive - to the OLR. Specifically, assuming a flat rotation 
curve, for the 4:1 ILR 
$\Omega_s=0.65\Omega_0$, for the 4:1 OLR $\Omega_s=1.35\Omega_0$, for the 
2:1 ILR $\Omega_s=0.3\Omega_0$, and for the 2:1 OLR $\Omega_s=1.7\Omega_0$. 
Corotation of each spiral pattern occurs at $\Omega_s=\Omega_0$. 

\subsection{Simulation method}

To investigate the effect of a second spiral density wave on the orbits of 
stars,
we ran simulations of test particles in a disk with two spiral density waves.
As done by \citet{desimone} the amplitude of each wave was varied,
however the pitch angle and pattern speed were held fixed during
the integrations. We are restricting this initial study to this type
of spiral arm growth (no swing amplifier mechanism).

\subsubsection{Initial conditions}

Positions of test particles (stars) were chosen randomly with a uniform
density distribution in both radial and azimuthal directions.
Thus, we distribute stars in an annulus of
an inner and outer radii $R_0-\Delta R$, $R_0+\Delta R$, where for most
of the runs $\Delta R$ is 0.3.  
All stars were given the same initial velocity $(u,v)=(0,0)$ consistent 
with a flat rotation curve and an initially cold stellar disk.  

We now describe the growth of the spiral density waves.
The amplitudes, $\epsilon_a,\epsilon_b$ are zero before $t_0$, grow linearly 
with time at $0<t<t_1$ and stay constant at $\epsilon_a=\epsilon_{max,a}$, 
$\epsilon_b=\epsilon_{max,b}$ after $t=t_1=4$ rotation periods. 

Simulations were run for 40 rotation periods, which corresponds to about
10 Gyr. Heating rates were estimated calculating the radial velocity dispersion 
(the standard deviation of the radial velocity component $u$), 
every rotational period. This was done for all stars in the 
annulus ($R_0-\Delta R$,$R_0+\Delta R$) where $\Delta R=0.05$.

The standard deviation was computed as the square root of the bias-corrected
variance:

\begin{equation}
\sigma_R = \sqrt{\sum_{i=1}^n (u_i-\bar{u})^2 \over {n-1}}
\end{equation}
where $n$ is the number of stars in the bin (the annulus of width $\Delta R=0.05$
centered on $R_0$, described above) and
$\bar{u}=(\sum_{i=1}^n u_i)/n$.
Table 1 shows the parameters for which simulations were run.

\subsubsection{Numerical accuracy}
Calculations were performed in double precision. We checked how energy  
was conserved in an unperturbed test run. The initial energy was compared to 
that calculated after 40 periods ($\approx 10$ Gyr) and the relative error 
was found to be $|\Delta E/E(0)|<8.48\times10^{-14}$. We also checked the 
conservation of the Jacobi integral $J$ in the presence of one periodic 
perturbation. The relative error found in this case was 
$|\Delta J/J(0)|<3.11\times10^{-12}$.

\section{One spiral pattern}
\label{sec:onepattern}

To tell the difference between the effect of one spiral pattern
and both patterns, we first ran comparison simulations with only one spiral 
pattern present. In each run we integrated 6000 particles. Parameters used for 
integrations are listed in Table \ref{table:all}.
\newline

\begin{figure*}
\epsscale{0.5}
\plotone{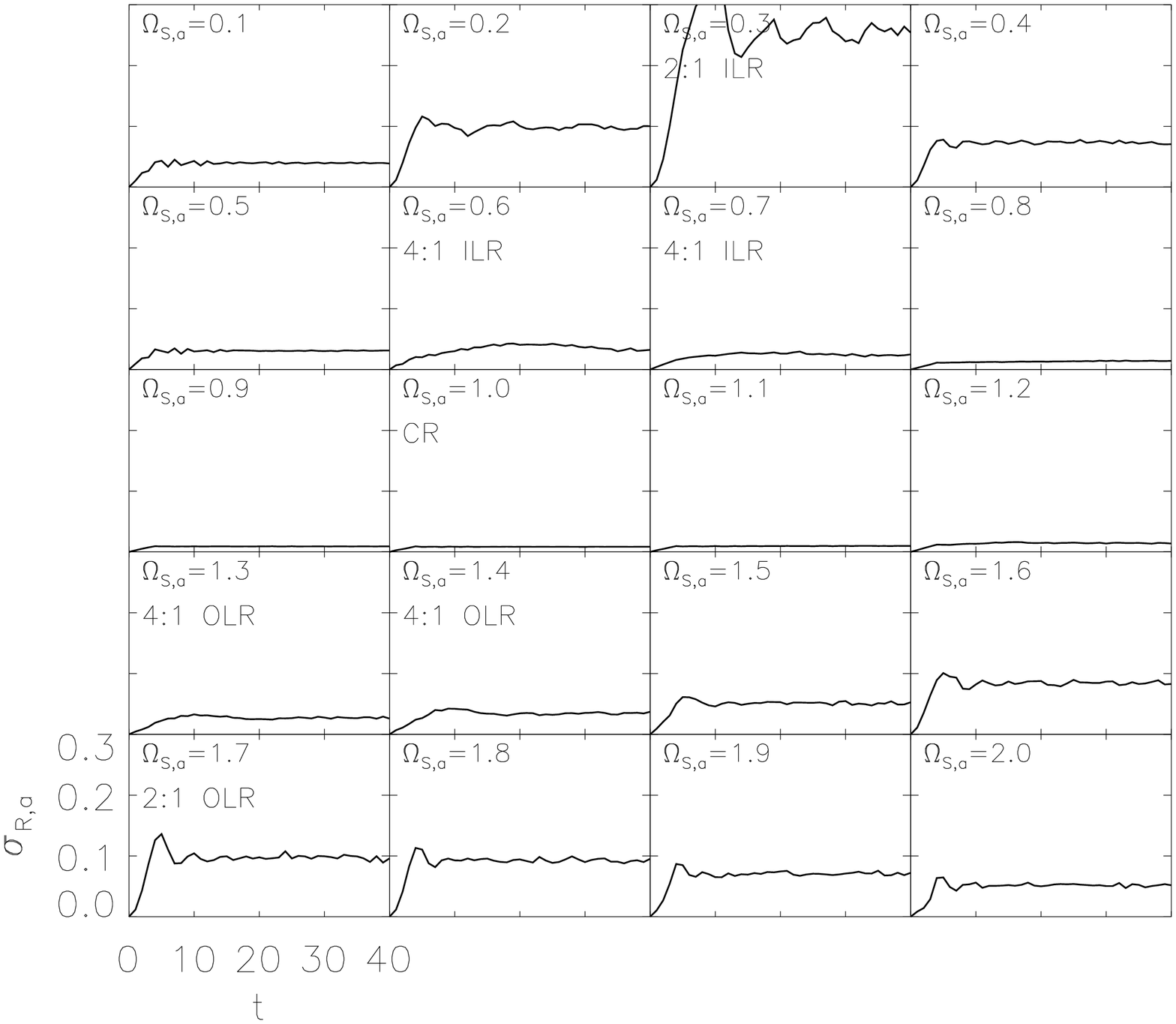}
\figcaption{
Radial velocity dispersion as a function of time for the primary 2-armed spiral 
pattern only. Each panel shows a different value of the angular velocity $\Omega_{s,a}$.
The gravitational potential amplitudes are $\epsilon_a = -0.006$ and 
$\epsilon_b = 0$. The 2:1 Inner Lindblad Resonance (ILR) and Outer Lindblad Resonance 
(OLR) occur at $\Omega_{s,a}=0.3\Omega_0$ and 
$\Omega_{s,a}=1.7\Omega_0$ respectively. Time is in units of rotational periods 
and the radial velocity dispersion $\sigma_R$ is in units of the circular velocity 
$V_0$. Note the low values of $\sigma_R$ around the corotation radius 
$\Omega_{s,a}=1.0$. There is no increase in $\sigma_R$ once the wave is grown.
\label{fig:just_e1}
}
\end{figure*}

In Fig.\ref{fig:just_e1} we plot the radial velocity dispersion, $\sigma_{R,a}$
as a function of time for the primary 2-armed spiral wave only. In all panels 
the amplitude of the gravitational perturbation is kept fixed at 
$\epsilon_a=-0.006$ once the wave is grown, and the pitch angle function is 
$\alpha_a=-6$ which corresponds to $p_a\approx-18 \degr$. Each panel in Fig. 
\ref{fig:just_e1} shows the increase of the radial velocity dispersion 
with time at a different pattern speed, $\Omega_{s,a}$.  

Logarithmic spirals are known to be self-similar. 
Due to this fact, changing the angular frequencies of the spiral waves
is equivalent to changing the radius.
Since we assumed a flat rotation curve, $V_0=1$ everywhere
and the angular velocity of stars varies with radius as $\Omega=V_0/R=1/R$. 
On the other hand, since the pattern speed, $\Omega_s$ does not vary with $R$,
i.e., the pitch angle is constant, a spiral rotates as a rigid body. 
It follows that in the units used in this paper we can just
substitute $R$ for $\Omega_{s,a}$ in Fig. \ref{fig:just_e1}, placing $R_0$ at 
$\Omega_{s,a}=1.0$. We note that since stars in the inner parts of a 
flat-rotation-curve galaxy complete one rotation period faster than stars 
in the outer parts, 40 rotation periods at $R_0$ translates as a larger
number of revolutions in the inner parts and a smaller number in the outer parts. 
Assuming a corotation radius near $R_0$ and a galactic radius of 
$\sim2R_0$, Fig. \ref{fig:just_e1} gives the heating as a function of radius
in the galaxy for one pattern.
The top left hand panel shows pattern speeds placing $R_0$ near
the 2:1 ILR. The lower right hand panel shows pattern speeds placing
$R_0$ near the 2:1 OLR. 

As the spiral amplitude, $\epsilon_a$, reaches its maximum value the 
radial velocity dispersion levels off.
As predicted by many previous theoretical investigations,
there is no heating after the wave is grown.
As expected, large radial velocity dispersion values are found 
only near the 2:1 Lindblad resonances ($\Omega_{s,a}=0.3, 1.7$). 
This is consistent with \citet{lb72} who showed that steady spirals heat 
stars only at the Lindblad resonances. Note that at and around corotation 
($\Omega_{s,a}=1.0$) random motions of stars were not increased.
\newline

\begin{figure*}
\epsscale{0.5}
\plotone{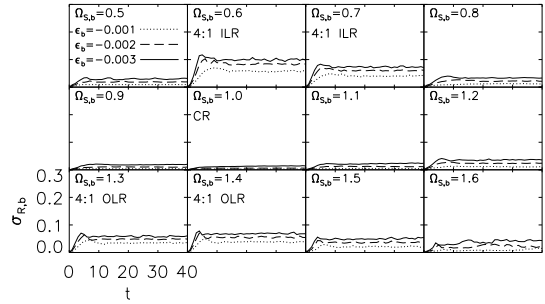}
\figcaption{
Radial velocity dispersion as a function of time for the secondary, 4-armed spiral
pattern only. Each panel shows a different value of the angular velocity 
$\Omega_{s,b}$. The 4:1 Inner Lindblad Resonance (ILR) and Outer Lindblad Resonance 
(OLR) occur at $\Omega_{s,b}$=0.35,1.65, respectively. Since we incremented 
$\Omega_{s,b}$ by 0.1 this figure does not show a panel exactly at OLR or ILR. Thus we
labeled the LRs on each side of their occurrence. Different line styles show 
different values of the spiral wave gravitational potential perturbation: 
$\epsilon_b = -0.001$ (dotted), $\epsilon_b = -0.002$ (dashed), and 
$\epsilon_b = -0.003$ (solid).
\label{fig:just_e2}.
}
\end{figure*}

The effect of a weak 4-armed spiral pattern on the radial velocity dispersion
of stars is shown in Fig. \ref{fig:just_e2} for 3 values of the gravitational 
potential perturbation amplitude, $\epsilon_b$.
Different line styles show $\epsilon_b=-0.001$ (dotted), $\epsilon_b=-0.002$
(dashed), and $\epsilon_b=-0.003$ (solid).
As in the case of the 2-armed spiral wave, relatively
large $\sigma_{R,b}$ values are observed only near the first order, 4:1 LRs 
($\Omega_{s,b}=0.65,1.35$).
Substituting $R$ for $\Omega_{s,b}$ 
in Fig. \ref{fig:just_e2} gives $\sigma_R$=$\sigma_R(R)$ in the range 
($R=0.5,R=1.6$). 
\newline

In both the 2- and 4-armed spiral structures the initial rapid increase ends
when the spiral amplitudes attain their maximum values
($t\approx4$ rotations) and are fixed for the rest of the runs.
This observation will later be contrasted with the effect of the 
simultaneous propagation of 2 spiral waves (see \S \ref{sec:t_evol_2sp}).

\section{Two spiral patterns grown simultaneously}

We next ran simulations with two spiral density waves moving at different 
pattern speeds. The spiral amplitudes were grown simultaneously 
over 4 rotation periods as in the case of only one spiral wave. 

\subsection{Heating in ($\Omega_{s,a}$, $\Omega_{s,b}$)-space}
\label{sec:contours}

\begin{figure*}
\epsscale{0.5}
\plotone{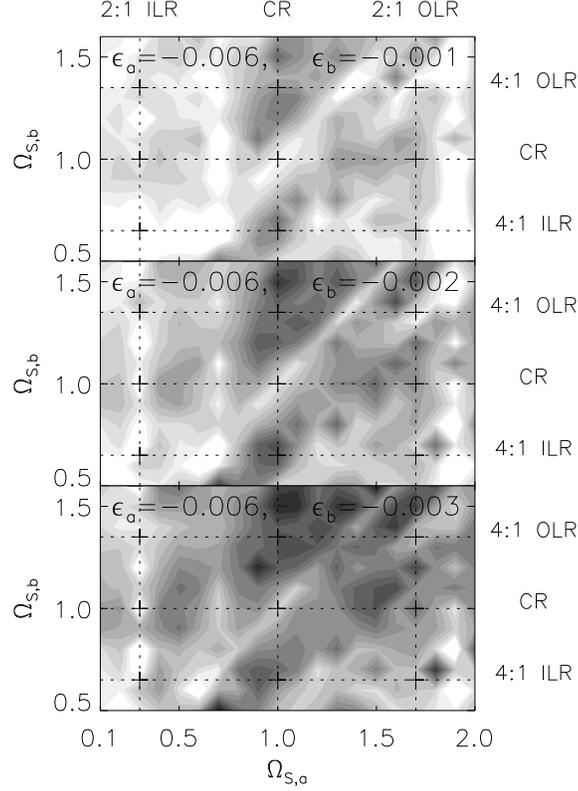}
\figcaption{
Contours show 
$\Delta\sigma_R\equiv\sigma_{R,ab}-\sqrt{{\sigma_{R,a}}^2+{\sigma_{R,b}}^2}$ 
for 3 values of the amplitude of the secondary spiral wave, 
$\epsilon_b=-0.001,-0.002,-0.003$, and a constant value of the the primary 
amplitude, $\epsilon_b=-0.006$ in ($\Omega_{s,a}$,$\Omega_{s,b}$)-space. 
Spiral amplitudes were grown simultaneously in 4 periods and were kept fixed 
after that. Here $\sigma_{R,a}$ and $\sigma_{R,b}$ are the 
radial velocity dispersions as a result of only the primary or 
only the secondary spiral patterns, respectively. These were time averaged over 
40 galactic rotations, excluding the time during which wave amplitudes grew
(4 rotation periods).
Similarly, $\sigma_{R,ab}$ is the time average of a simulation run with both 
patterns perturbing the test particles. Consequently, $\Delta\sigma_R$ gives 
the additional heating caused by the combined effect of the two perturbers as 
opposed to the sum of the heating that would have been produces if each spiral 
wave propagated alone.
In all panels $\Delta\sigma_R$ was normalized to its maximum value in the 
three panels allowing a proper comparison among the panels.
The pattern speed of the primary wave, $\Omega_{s,a}$, varies between 0.1 
and 2.0 and that of the secondary one, $\Omega_{s,b}$, varies between 0.5 
and 1.6 in units of the angular rotation rate, $\Omega_0=1$ at $R_0=1$.
The dashed lines show the locations of corotation and Lindblad resonances due 
to each of the two spiral wave perturbations. Next to each dashed line we 
labeled the resonances by denoting the corotation resonance with "CR", and 
the outer and inner Lindblad resonances by "OLR" and "ILR".
For a given set of pattern speeds, a line at 45$\degr$ slope predicts the 
variation of $\Delta\sigma_R$ with radius,
$R$, where the intercept of the line is set by the relative angular frequency,
$\Delta\Omega_s\equiv\Omega_{s,a}-\Omega_{s,b}$.
Large values of $\Delta\sigma_R$ are primarily observed along the lines 
$\Omega_{s,a}=1.0$ and $\Omega_{s,b}=1.0$, which
correspond to the CR of each pattern, with the exception of the intersection 
of these lines; $\Delta\sigma_R(R)$ is a strong function of radius.
\label{fig:eps12_heat}
}
\end{figure*}

First we carried out a set of integrations with 1000 particles going through all 
the possible combinations of the two pattern speeds in the range (0.1,2.0) for
the $\Omega_{s,a}$ and (1.5,1.6) for $\Omega_{s,b}$, in steps of 0.1, and in 
units of the angular rotation rate, $\Omega_0$ at $R_0$. 
The chosen ranges of 
$\Omega_{s,a}$ and $\Omega_{s,b}$ ensure that we cover the Lindblad 
and corotation resonances of each spiral wave. The results are shown as the 
($\Omega_{s,a}$, $\Omega_{s,b}$)-space contour plots in Figures 
\ref{fig:eps12_heat} and \ref{fig:eps12_heat_only}, described below. 

Horizontal axis is $\Omega_{s,a}$ and vertical axis is $\Omega_{s,b}$ for 
each panel. Each panel corresponds to a different secondary spiral wave 
amplitude, $\epsilon_b=-0.001,-0.002,-0.003$. The primary pattern amplitude 
is constant for all panels at $\epsilon_a=0.6$. Spiral amplitudes were grown 
simultaneously in 4 rotational periods and were kept fixed after that. 

In Fig. \ref{fig:eps12_heat} contours show 
\begin{equation}
\label{eq:delta_sigma}
\Delta\sigma_R\equiv\sigma_{R,ab}-\sqrt{{\sigma_{R,a}}^2+{\sigma_{R,b}}^2}. 
\end{equation}
Here $\sigma_{R,a}$
and $\sigma_{R,b}$ are the radial velocity dispersions as a result of only 
the primary or only the secondary spiral patterns, respectively. These were 
time averaged over 40 galactic rotations, excluding the time during which wave 
amplitudes grew (initial 4 periods).
Similarly, $\sigma_{R,ab}$ is the time average of a simulation run with both 
patterns perturbing the test particles. Consequently, $\Delta\sigma_R$ gives 
the additional heating caused by the combined effect of the two perturbers as
opposed to the sum of the heating that would have been produced if each spiral 
wave propagated alone. 

The dashed lines show the locations of corotation and Lindblad resonances due 
to each of the two spiral wave perturbations. Next to each dashed line we 
labeled the corresponding resonances by denoting the corotation resonance with
"CR", and the outer and inner Lindblad resonances with "OLR" and "ILR". 

Grayscale for $\Delta\sigma_R$ was normalized to the maximum value of the 
three panels allowing a proper comparison among the panels. Darker regions 
corresponds to larger values of $\Delta\sigma_R$. 

We see that $\Delta\sigma_R>0$ for most combinations of angular velocities 
which implies that spiral waves interfere and produce additional heating. 
In all the panels maximum values of $\Delta\sigma_R$ 
are attained mainly along the lines $\Omega_{s,a}$=1.0 and $\Omega_{s,b}$=1.0 
corresponding to the CR of each spiral wave. A notable exception to this trend
is the intersection of these lines where patterns do not have relative motion 
with respect to each other. As the absolute value of the secondary spiral wave 
amplitude, $\vert\epsilon_b\vert$, is increased from 0.001 to 0.003 structure 
in the $\Omega_{s,a}$,$\Omega_{s,b}$-plane seems not to change significantly, 
however $\Delta\sigma_R$ is increased. 

It is interesting to note the lack of heating due to the effect of the coupling 
between the two spiral waves at the 2:1 ILR and on the outside of the
2:1 OLR of the 2-armed spiral wave, and at the vertical line $\Omega_a\approx0.7$
which is just outside the 4:1 ILR with the same perturbation.

We mentioned earlier that changing $\Omega_s$ and keeping $R$ constant
is equivalent to changing $R$ and keeping $\Omega_s$ constant. 
Thus, for a given set of pattern speeds, a line at 45$\degr$ slope in 
Fig. \ref{fig:eps12_heat} predicts the variation of $\Delta\sigma_R$ with radius,
$R$, where the intercept of the line is set by the relative angular frequency, 
$\Delta\Omega_s\equiv\Omega_{s,a}-\Omega_{s,b}$. Depending on the given 
$\Delta\Omega_s$, a galaxy heated by two spiral density waves would exhibit 
different variation of the radial velocity dispersion with radius. 
For example, if $\Delta\Omega\approx0$ then the galaxy would not be heated 
efficiently. This is expected since, as mentioned earlier, steady 
spirals heat stars only at the LRs. If, on the other hand, 
$\Delta\Omega\approx0.3$ then the currently examined heating mechanism would 
prove quite efficient and would effect stars over a wide region. It is evident 
from the figure that heating takes place 
mainly near the CR and OLR of primary, and near the CR of the secondary pattern.
\newline

Similarly to the $\Delta\sigma_R$ contours shown in Fig. \ref{fig:eps12_heat}
we plot contours of the radial velocity dispersion as the result of the 
simultaneous propagation of the two spiral waves, $\sigma_{R,ab}$, in 
Fig. \ref{fig:eps12_heat_only}. 

Here maximum values 
are attained at the 2:1 ILR of the primary, $\epsilon_a=-0.006$ spiral wave. 
We believe that in reality the 2:1 ILR is too strong to maintain spiral 
structure for strong spirals so we regard the results observed in 
this location on the $\Omega_{s,a}$, $\Omega_{s,b}$-plane as unphysical.
It is possible that in reality spiral amplitude decreases as this resonance is
reached or that spiral arms end there.
Aside from the 2:1 ILR of the 2-armed spiral wave in all the panels high
values of $\sigma_{R,ab}$ are observed around the 2:1 OLR of the primary spiral
wave or near resonances with the primary and the secondary pattern taking 
place at the same time (where dashed lines cross).
As in Fig. \ref{fig:eps12_heat} an exception to the last statement is the point 
($\Omega_{s,a}=1.0$, $\Omega_{s,b}=1.0$) for which the lowest values in 
$\sigma_{R,ab}$ are reached. 
Again, this corresponds to both patterns 
placing stars at the CR and thus no relative motion between them.

In the following section we chose interesting points corresponding to
specific combination of the pattern angular velocities in Fig.
\ref{fig:eps12_heat} and plotted the radial velocity dispersion, 
$\sigma_{R,ab}$, as function of time, $t$, to compare to the effect of the 
single spiral waves described in \S \ref{sec:onepattern}. Parameters for all 
runs can be found in Table \ref{table:all}.

\subsection{Time evolution of the radial velocity dispersion
at the CR and OLR of primary pattern}
\label{sec:t_evol_2sp}

\begin{figure*}
\epsscale{0.5}
\plotone{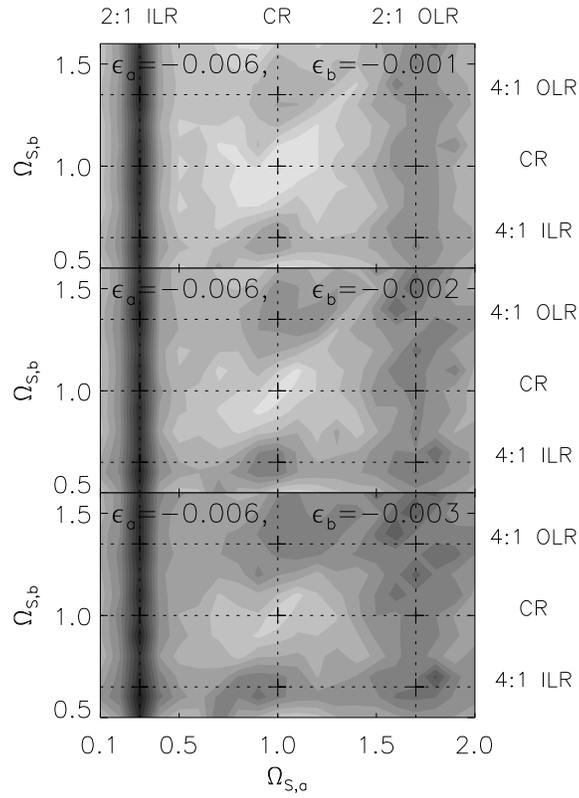}
\figcaption{
Contour plots of the heating produced by the propagation of 2 spiral waves.
Identical setup as in Fig. \ref{fig:eps12_heat} but contours show 
$\sigma_{R,ab}$, instead of $\Delta\sigma_R$. Aside from the 2:1 LRs of the 
primary, 2-armed spiral wave, maximum values of $\sigma_{R,ab}$ are found
at and around the CR with the primary and 4:1 LRs of the secondary patterns.
As in Fig. \ref{fig:eps12_heat}, at the CR with both spiral waves very low 
radial velocity dispersion is seen (here it is a minimum). For all panels
$\sigma_{R,ab}$ was normalized to the largest value in the figure. 
\label{fig:eps12_heat_only}
}
\end{figure*}

\begin{figure*}
\epsscale{0.5}
\plotone{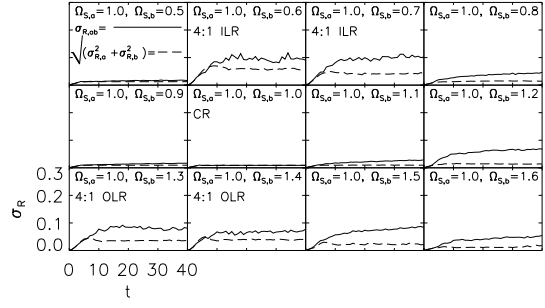}
\figcaption{
Time evolution of the radial velocity dispersion, $\sigma_{R,ab}$, as a result 
of the simultaneous propagation of a 2- and a 4-armed 
spiral patterns (solid line). Time is in units of galactic rotations.
The 2-armed spiral structure has the same parameters as the panel in
Fig.\ref{fig:just_e1} for which $\Omega_{s,a}=1.0$ (CR), whereas the 4-armed
spiral wave has different angular velocity in each panel. The time evolution
of the radial velocity dispersion due to the latter perturbation alone, 
$\sigma_{R,b}$, was shown in Fig. \ref{fig:just_e2}. 
For comparison, in addition to $\sigma_{R,ab}$ (solid line) we also plotted
$\sqrt{{\sigma_{R,a}}^2+{\sigma_{R,b}}^2}$ (dashed line) 
(see Eq. \ref{eq:delta_sigma}).
The values of the perturbation amplitudes are $\epsilon_a = -0.006$ and 
$\epsilon_b = -0.001$. We note the following dynamical effects around the 4:1 
ILR and OLR in the case of $\sigma_{R,ab}$:  
(1) $\sigma_{R,ab}$ increases up to a factor of $\approx3$  compared to the 
dispersion expected from the noninteracting spiral waves, 
$\sqrt{{\sigma_{R,a}}^2+{\sigma_{R,b}}^2}$;
(2) $\sigma_{R,ab}$ grows with time;
(3) resonances have effect over a larger range of angular velocities,
compared to $\sqrt{{\sigma_{R,a}}^2+{\sigma_{R,b}}^2}$, and
thus over a larger range of radii.
\label{fig:both_1_o10}
}
\end{figure*}

\begin{figure*}
\epsscale{0.5}
\plotone{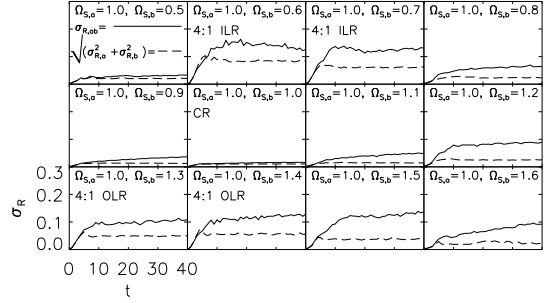}
\figcaption{
Same as Fig.\ref{fig:both_1_o10} but with the secondary gravitational potential 
amplitude $\epsilon_b = -0.002$. 
The radial velocity dispersion at the LRs saturates faster than that in Fig. 
\ref{fig:both_1_o10} as $\vert\epsilon_b\vert$ was increased from 0.001 to 
0.002.
\label{fig:both_2_o10}
}
\end{figure*}

\begin{figure*}
\epsscale{0.5}
\plotone{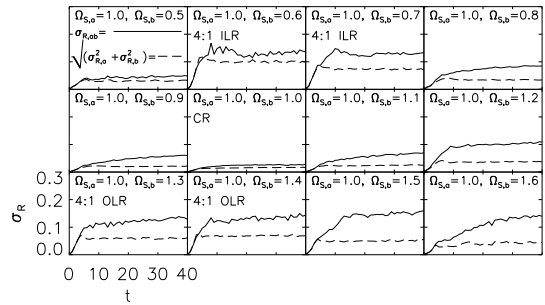}
\figcaption{
Same as Figures \ref{fig:both_1_o10} and \ref{fig:both_2_o10} but with 
the secondary gravitational potential amplitude $\epsilon_b = -0.003$. Note
the further decrease in the saturation time of $\sigma_{R,ab}$ as compared to 
Figures \ref{fig:both_1_o10} and \ref{fig:both_2_o10}.
\label{fig:both_3_o10}
}
\end{figure*}

To better understand the nature of the process giving rise to the structure 
seen in Figures \ref{fig:eps12_heat} and \ref{fig:eps12_heat_only} we plotted 
the time evolution of $\sigma_{R,ab}$ and 
$\sqrt{{\sigma_{R,a}}^2+{\sigma_{R,b}}^2}$ 
(as defined in the previous subsection, see Eq. \ref{eq:delta_sigma}). 
These dispersions are shown in 
Figures \ref{fig:both_1_o10}-\ref{fig:both_3_o10} as functions of time. 
We chose to consider stars near the corotation resonance with the primary 
pattern and with different values for the secondary pattern speed.
For reference, the time average values of $\Delta\sigma_R$ and $\sigma_{R,ab}$ 
are given by the vertical line $\Omega_{s,a}=1.0$ in Fig. \ref{fig:eps12_heat} 
and the same line in Fig. \ref{fig:eps12_heat_only}, respectively.

The solid lines in Figures \ref{fig:both_1_o10}-\ref{fig:both_3_o10} depict 
the effect of the 
simultaneous propagation of a 2- and a 4-armed spiral patterns on the 
radial velocity dispersion, $\sigma_{R,ab}$, as a function of time, $t$, in 
units of galactic rotation periods. Figures \ref{fig:both_1_o10}, 
\ref{fig:both_2_o10}, and \ref{fig:both_3_o10} differ only in the value of the 
amplitude of the secondary spiral wave, to wit, 
$\epsilon_b=-0.001,-0.002,-0.003$, respectively. 
The value of the primary spiral wave amplitude is fixed for 
all 3 Figures with $\epsilon_{a}=-0.006$.
The 2-armed spiral structure has the same parameters as that shown in 
Fig.\ref{fig:just_e1} for which $\Omega_{s,a}=1.0$ (CR), whereas the 4-armed 
spiral wave has different angular velocity in each panel of Figures 
\ref{fig:both_1_o10}-\ref{fig:both_3_o10}. The
dispersion due to the latter perturbation alone, $\sigma_{R,b}$, 
was shown in Fig. \ref{fig:just_e2}. 
For comparison, in addition to $\sigma_{R,ab}$ (solid line) we also plotted 
$\sqrt{{\sigma_{R,a}}^2+{\sigma_{R,b}}^2}$ (dashed line) (see Eq. 
\ref{eq:delta_sigma}). 

The 4:1 resonances of the secondary spiral wave are seen in the lower left 2 
panels (4:1 OLR) and the upper middle 2 panels (4:1 ILR).
Note the big dynamical changes at those locations for the case of $\sigma_{R,ab}$:
\newline
(1) $\sigma_{R,ab}$ increases up to a factor of $\approx$ 3 compared to the 
dispersion expected from the noninteracting spiral waves, 
$\sqrt{{\sigma_{R,a}}^2+{\sigma_{R,b}}^2}$ (i.e., panels with 
$\Omega_b=1.2,1.5$, Figures \ref{fig:both_1_o10}-\ref{fig:both_3_o10});
\newline
(2) $\sigma_{R,ab}$ grows with time. In Fig. \ref{fig:both_1_o10} at 
$\Omega_{s,b}=0.6,0.7,1.4$ $\sigma_{R,ab}$ increases until $t\approx10$; for 
$\Omega_{s,b}=0.8,1.2,1.5,1.6$ $\sigma_{R,ab}$ increases at even later times 
of 20-30 periods. This does not happen when only one pattern is present.
\newline
(3) resonances have effect over a larger range of angular velocities,
compared to $\sqrt{{\sigma_{R,a}}^2+{\sigma_{R,b}}^2}$, and
thus over a larger range of radii.

As noted earlier, in the case of only one spiral 
density wave large effects are primarily observed near the 2:1 Lindblad 
resonances for the 2-armed, and near the 4:1 Lindblad resonances for 
the 4-armed spiral patterns during the growth of the perturbations only. 
In contrast, when two spiral waves at a particular combination of pattern 
speeds are present, $\sigma_{R,ab}$ keeps increasing even after the maximum 
values of the potential amplitudes, $\epsilon_a$ and $\epsilon_b$, are reached. 
We expect the most likely explanation for this is that
the particular combination of pattern speeds, namely, a 2-armed spiral at
the CR and a 4-armed one near the ILR or OLR, creates
stochastic motion due to resonance overlap. This is analogous to the 
bar - spiral wave overlap described by \citet{quillen_chaos}, found to produce
a stochastic region in the vicinity of the Galactic bar's OLR. In these 
conditions stars are expected to diffuse radially in real space 
(see \S \ref{sec:orbits_2sp}).

The dispersion $\sigma_{R,ab}$ grows with time up to a specific value of $t$ 
(saturation time) after which it does not increase anymore. 
It is interesting to note the decrease in the saturation time as the 
strength of the secondary spiral increases in Figures 
\ref{fig:both_1_o10}-\ref{fig:both_3_o10}. 

If the observed increase of $\sigma_{R,ab}$ with $t$ is due to the presence of 
both the CR of the primary, 2-armed and the LRs of the secondary, 4-armed spiral
waves, then one would expect that as one moves away from the LRs of the secondary 
pattern the effect on the radial dispersion would diminish. 
This is generally the trend observed in Figures 
\ref{fig:both_1_o10}-\ref{fig:both_3_o10}, however on both sides of the 4:1 OLR
the opposite effect is seen. 
At $\Omega_{s,b}=1.2,1.5$ large values of $\sigma_{R,ab}$ are still evident. 
In fact, at the points ($\Omega_{s,a},\Omega_{s,b},t$)=(1,1.2,40),(1,1.5,40)
in Fig. \ref{fig:both_1_o10} 
the ratio $\sigma_{R,ab}/\sqrt{{\sigma_{R,a}}^2+{\sigma_{R,b}}^2}$ 
attains its maximum value of $\approx3.5$. These high values of $\Delta\sigma_R$ 
on each side of the 4:1 OLR of the 4-armed pattern were anticipated, considering
the structure along the line $\Omega_{s,a}=1.0$ in Fig. \ref{fig:eps12_heat}. 
\newline

In Fig. \ref{fig:both_3_o17} we plot the time evolution of $\sigma_{R,ab}$ and
$\sqrt{{\sigma_{R,a}}^2+{\sigma_{R,b}}^2}$ for $\Omega_{s,a}=1.7$.
At this angular velocity of the primary pattern 
particles are placed at the 2:1 OLR. The gravitational perturbation amplitude
of the secondary wave is $\epsilon_b=-0.003$; the primary one has the 
default value of $\epsilon_a=-0.006$. Obviously, the fraction 
$\sigma_{R,ab}/\sqrt{{\sigma_{R,a}}^2+{\sigma_{R,b}}^2}$ is much smaller than
that observed in Figures \ref{fig:both_1_o10}-\ref{fig:both_3_o10}. 
This was expected in view of Figures 
\ref{fig:eps12_heat} and \ref{fig:eps12_heat_only}.
\newline

It is clear from Figures \ref{fig:both_1_o10}-\ref{fig:both_3_o17} that the 
velocity dispersion in the presence of 2 patterns, $\sigma_{R,ab}$, exceeds that
expected from each individual pattern $\sqrt{{\sigma_{R,a}}^2+{\sigma_{R,b}}^2}$.
Furthermore, $\sigma_{R,ab}$ continues to increase with time for some 
values of $\Omega_{s,a},\Omega_{s,b}$. This contrasts with the predictions for
heating due to transient spiral density waves. Additional heating occurs in the
presence of 2 waves and the interplay of the 2 waves contributes to additional
heating after they have ceased to grow.
\newline

We also considered the more realistic situation in which the second 
spiral density wave was grown at a later time than the first one.
At all parameters kept the same as those used to produce Figures 
\ref{fig:both_1_o10}, \ref{fig:both_2_o10}, and \ref{fig:both_3_o10},
we ran simulations in which the starting time for 
the secondary spiral wave amplitude was set to $2, 4, 7, 15$, and $20$.
We found that the sole effect of the delayed second
spiral was to delay the starting point of the radial dispersion increase. 

\section{Orbits of stars near resonances in the rotation frame of the 
pattern}
\label{sec:orbits}

To understand what causes the heating observed in Figures 
\ref{fig:both_1_o10}-\ref{fig:both_3_o17} 
we will next look at the shapes of the orbits of test particles when subjected 
to only one of the 2 spiral perturbations and then compare those to the 
orbits of particles in the presence of both.

The description of the particle trajectories below
is based on movies we made for each simulation run in the current section, 
\ref{sec:orbits}. Snaps from the movies are shown in the figures below.

\subsection{Orbits of stars subjected to one spiral wave only}

We offer a qualitative description of the orbits of stars near corotation 
with the primary spiral wave only, as seen in a reference frame rotating with
the spiral pattern.
To simplify this task as much as possible we integrated 1000 particles
placed at different radius $R$ and uniformly distributed azimuthally. Different 
panels in Fig. \ref{fig:m2_CR} show particles initially moving:
\newline
(a) just inside the CR ($R=0.9$), 
\newline
(b) exactly at the CR ($R=1.0$), and
\newline
(c) just outside the CR ($R=1.1$). 
These panels are shown for $t=40$ rotation periods for 3 separate simulations.
Dashed circles show the initial distribution of the particles.

Just inside of the corotation circle stars are moving faster than the spiral 
pattern (Fig.\ref{fig:m2_CR}, a). In a reference frame rotating with the spiral, 
their orbits are confined, depending on the azimuthal angle of each star at the 
time the spiral perturbation is turned on, to one of 3 types of orbits.
These are an oval orbit and two banana-shaped closed curves situated outside 
the initial circle, the cusps of which coincide with the two spiral arms.

Just outside of the corotation radius (Fig.\ref{fig:m2_CR}, c) stars follow 
similar orbits to the ones described above, except that because the spiral wave 
moves faster than the stars, the "banana" orbits are situated inside both the 
oval-shaped orbit and the initial circle (at $R_0$). Furthermore, the oval orbit 
is flattened at a $90\degr$ phase difference compared to the same type of orbit 
inside of the CR.

Finally, exactly at corotation a more complicated situation is
observed (Fig.\ref{fig:m2_CR}, b). As seen in the reference frame of the spiral,
stars move on 2 families of nested closed curves similar to the 2 "banana" 
orbits discussed above. Each such orbit protrudes equally radially inward and 
outward. Note, that the third, oval shaped orbit is not present here. Stars 
which happened to be at the arms at the time the wave was turned on, stayed 
there and the rest are not allowed to cross an arm at any time.

We talked about the orbits of stars but what Fig. \ref{fig:m2_CR} actually
shows are snapshots at a single time, $t=40$ rotations. 
During the entire simulation a given closed contour on which stars move does not
change shape or orientation as seen in the reference frame of the spiral wave. 
Consequently, a plot showing the orbital trajectory
of a single particle would look exactly like the snapshots in these pictures.
However, if a contour on which stars move {\it does} change shape or orientation
then it cannot be regarded as the trajectory that a single star would follow.
\newline

Similarly to the CR of the 2-armed spiral wave, information about the orbits of
stars placed at and near the 4:1 OLR ($\Omega_b=1.65$) of the secondary, 4-armed 
spiral wave can be obtained from Fig. \ref{fig:m4_ILR}. In all panels 
all parameters are kept the same except for the radius at which stars were 
initially distributed. The pattern amplitudes are $\epsilon_b=-0.002$, 
$\epsilon_a=0$ and angular velocity is $\Omega_b=1.65$.
Panels show stars initially moving on a circle of radius:
\newline
$R=0.9$ or just inside the 4:1 ILR (Fig. \ref{fig:m4_ILR}, a), 
\newline
$R=1.0$ or exactly at the 4:1 ILR (Fig. \ref{fig:m4_ILR}, b), and
\newline
$R=1.1$ or just outside the 4:1 ILR (Fig. \ref{fig:m4_ILR}, c). 

The snapshots are at time $t=40$ galactic rotations.
As we discussed at the end of the previous paragraph, the curve on which test
particles lie at a point in time is not necessarily the trajectory which a 
single particle would follow in time (in a frame rotating with the pattern). It 
turns out that only Fig. \ref{fig:m4_ILR}, (b) depicts a curve which changes 
shape and orientation as the system develops in time. Therefore, we do not 
regard Fig. \ref{fig:m4_ILR}, (b) as the trajectory of a particle.

The particle trajectories (in a frame rotating with the pattern) in panels 
(a),(c) of Fig. \ref{fig:m4_ILR} form as 
soon as the spiral wave amplitude, $\epsilon_b$, is grown (4 galactic rotations) 
and do not change until the end of our runs (40 rotation periods). However, 
exactly at the 4:1 ILR (Fig. \ref{fig:m4_ILR}, b) the rectangular curve starts 
to form at $t\approx2$ rotations 
with its vertices supporting the spiral structure and starts to spin clockwise 
in the rotation frame of the pattern. As it does so its vertices sharpen to 
achieve the curve seen in Fig. \ref{fig:m4_ILR}, (b). The curve goes back
to being oblate as the rectangle's vertices reach the arms again. This process
is repeated with a period of about 9 galactic rotations. 
Note that star orbits support the spiral structure outside the 4:1 ILR 
(Fig.\ref{fig:m4_ILR}, c) but are out of phase with the perturbation inside 
of it (Fig.\ref{fig:m4_ILR}, a). The orientation of the orbits changes at the
resonance. It is known that strong 2-armed density perturbations can excite 
square-shaped orbits as well \citep{cont86,q+m}. However, in this case (secondary 
order in the epicyclic amplitude 4:1 ILR) the spirals are supported by star orbits 
interior to the 4:1 ILR and are not supported by the orbits of stars having their 
guiding radii exterior to it. This behavior prompted \cite{cont85} to propose that
2-armed spiral waves ended at the 4:1 ILR.
\newline

To complete the picture we lastly show the behavior of stars when subjected to
the 4:1 OLR of the secondary, 4-armed spiral wave only. Fig. \ref{fig:m4_OLR}
shows a snapshot of the real-space particle distributions at time $t$=40 galactic
rotations. Pattern speeds and gravitational potential amplitudes were the same 
for each panel, $\Omega_b=1.35$ and $\epsilon_b=0.002$, respectively. The initial 
distribution of each simulation was at:
\newline
$R=0.9$ or just inside the 4:1 OLR (Fig. \ref{fig:m4_ILR}, a), 
\newline
$R=1.0$ or exactly at the 4:1 OLR (Fig. \ref{fig:m4_ILR}, b), and
\newline
$R=1.1$ just outside the 4:1 OLR (Fig. \ref{fig:m4_ILR}, c). 

As in Fig. \ref{fig:m4_ILR}, inside and outside the resonance the initially 
circular curve did not change shape or orientation, thus it represents the 
approximately closed orbit of stars in a frame rotating with the pattern 
(Fig. \ref{fig:m4_ILR}, a, c). On the other hand, Fig. \ref{fig:m4_ILR}, (b) 
shows a curve which spins as the system develops in time. 
\newline

In all the cases but Figures \ref{fig:m4_ILR}, (b) and \ref{fig:m4_OLR}, (b), the 
initial circles assumed some particular shape at the end of the growth of the 
spiral waves and did not change for the remaining time. 
To check whether this was the result of an adiabatic perturbation we also tried 
using a sudden perturbation by growing the spiral 
in just 0.1 period, as opposed to the default value of 4 periods, and found 
similar results.

\subsection{Orbits of stars subjected to 2 spiral density waves moving at
different angular velocities} 
\label{sec:orbits_2sp}

\begin{figure*}
\epsscale{0.5}
\plotone{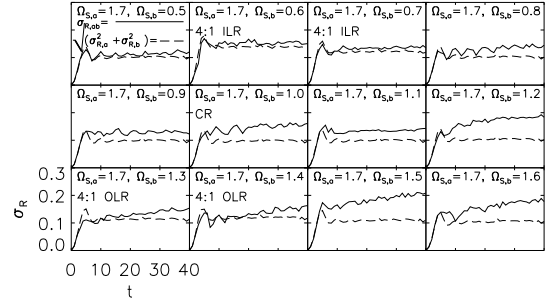}
\figcaption{
Same as Figure \ref{fig:both_3_o10} 
but with the primary spiral wave placing test particles at the 2:1 OLR
($\Omega_{s,a}=1.7$), instead of the CR. As in Figures 
\ref{fig:both_1_o10}-\ref{fig:both_3_o10}, both spirals were grown simultaneously 
in the initial 4 rotation periods; the perturbation amplitudes are 
$\epsilon_a = -0.006$ and $\epsilon_b = -0.003$.
\label{fig:both_3_o17}
}
\end{figure*}

In this section we attempt to look for an answer to the following question:
What process gives rise to the continuous increase of the radial 
velocity dispersion, $\sigma_R$, with time, $t$, in Figures 
\ref{fig:both_1_o10}-\ref{fig:both_3_o17}?

We will only look at the following combinations of patterns:
\newline
(i) a 2-armed spiral wave at the CR ($\Omega_{s,a}=1.0$) propagating 
simultaneously with a 4-armed one just outside the 4:1 ILR ($\Omega_{s,b}=0.7$).
\newline
(ii) a 2-armed spiral wave at the CR ($\Omega_{s,a}=1.0$) propagating 
simultaneously with a 4-armed one just inside the 4:1 OLR ($\Omega_{s,b}=1.2$). 

The sole effect of a 2-armed spiral wave, moving at a pattern speed which 
places stars at the CR, on the stars' orbits was shown in panel (b), Figure 
\ref{fig:m2_CR}. 
The trajectory of a star moving exterior to the 4:1 ILR ($\Omega_{s,b}=0.7$,
considered in (i) above) is closest to the curve seen in panel (c), Fig. 
\ref{fig:m4_ILR}, which could be perceived as the result of a simulation with 
an initial spatial distribution on a circle of radius $R=1.0$ and a pattern 
speed of $\Omega_{s,b}=0.75$ (see 3-rd paragraph of \S \ref{sec:onepattern}). 
Similarly, the trajectory of a star moving inside the 4:1 OLR ($\Omega_{s,b}=1.2$,
considered in case (ii) above) is closest to the curve seen in 
panel (a), Fig. \ref{fig:m4_OLR} which corresponds to an initial spatial 
distribution on a circle of radius $R=1.0$ and a pattern
speed of $\Omega_{s,b}=1.25$. 

Remember that in Figures \ref{fig:both_1_o10}-\ref{fig:both_3_o17} initially we 
distributed stars in the annulus 
($R_0-\Delta R$, $R_0+\Delta R$) with $\Delta R=0.3$ and $R_0=1$. 
Trajectories of test particles were followed for 40 rotations 
(the lifetime of the Galaxy at $R_0$) and $\sigma_R$ was computed every period 
in the annulus ($R_0-\Delta R$, $R_0+\Delta R$) with $\Delta R=0.05$. 
Because of the differential
rotation of stars in the disk, the initial velocities of all test particles are
fixed at $(v,u)=(V_0,0)$ so it takes longer for particles at larger $R$ to 
complete one period compared to particles at smaller $R$. 
On the other hand, spirals move as rigid bodies.
It is obvious that particles moving at different radii would see the spiral 
structure moving at different pattern speed. Stars contributing to
$\sigma_R$ could have come from various radii, thus responding to different
pattern-speed spiral waves. This is a complicated process.
To simplify it, considering first the panel for which $\Omega_{s,b}=0.7$ in
Fig. \ref{fig:both_2_o10} (case (i) above), we give the following illustration:

(1) One can imagine the initial spatial distribution of particles
is made of rings of stars at different radii between $R_0-\Delta R$ and 
$R_0+\Delta R$. Thus one can simplify the problem by looking at rings of stars.
On a particular ring all stars feel a spiral pattern moving at the same 
pattern speed. Furthermore, a plot of the time development of the spatial 
distribution of stars would be much more informative than a similar plot of a 
simulation run with the set up used for Fig. \ref{fig:both_2_o10}, panel with 
$\Omega_{s,b}=0.7$. 

(2) Integrate 3 such rings corresponding to specific radii in the 
initial annulus distribution of Fig. \ref{fig:both_2_o10}. We chose those
to be at $R=0.9,1.0,1.1$. 

(3) Plot the spatial distribution of stars at different times to see 
whether particles, initially positioned on the circle of radius $R$, diffuse 
radially as was speculated earlier.  
\newline

Plotted in Fig. \ref{fig:m2m4oCRo07} are 3 steps of the 40-rotation-period time 
evolution of 3 rings of stars at the radii chosen in step 2 above.
Rows from top to bottom show particles subject to 2 pattern 
speeds with angular velocities $\Omega_{s,a}=1.0$, $\Omega_{s,b}=0.7$ 
(case (i) above) and initially positioned in a ring of radius: 
\newline
$R=0.9$ (Fig. \ref{fig:m2m4oCRo07}, a),
\newline
$R=1.0$ (Fig. \ref{fig:m2m4oCRo07}, b), and
\newline
$R=1.1$ (Fig. \ref{fig:m2m4oCRo07}, c).

Snapshots are at times $t=10,25$, and 40 galactic rotation periods. The crosses 
show the position of the galactic center. 

The 3 systems in Fig. \ref{fig:m2m4oCRo07} behave very differently compared to 
the ones depicted in Figures \ref{fig:m2_CR} and \ref{fig:m4_ILR}. 
Note that stars {\it do} diffuse radially. Orbits are not confined to narrow regions
in phase space as was observed in Figures \ref{fig:m2_CR}-\ref{fig:m4_OLR}.
We see very different dynamics in the trajectories of particles. 
In the first snap (at $t$=10) it is clear that inside 
the CR stars are scattered radially outward whereas the opposite effect is seen
exterior to the CR. This is similar to the behavior of stars in Fig. 
\ref{fig:m2_CR} and thus must be related to the proximity of the CR. However, 
stars do not lie on a specific curve as in the case of only one perturbation. 
As one moves to panels at $t=25,40$ a randomization in the azimuthal direction is 
apparent. As was seen in Fig. \ref{fig:both_2_o10}, panel with $\Omega_{s,b}=0.7$,
saturation time was reached at around $t\approx10$ periods. This is consistent 
with Fig. \ref{fig:m2m4oCRo07} where the real space particle distributions in the 
panels with $t=25,40$ for each row do not appear much different. 
Note that, in a simplified way, the time development of the real-space 
distribution of stars in Fig. \ref{fig:both_2_o10}, panel with $\Omega_{s,b}=0.7$,
can be obtained by overlapping the 3 rows in Fig. \ref{fig:m2m4oCRo07}. 
\newline

Similarly to Fig. \ref{fig:m2m4oCRo07}, radial diffusion of stars is expected 
in the vicinity of the CR due to the 2-armed pattern and the 4:1 OLR of the 
secondary, 4-armed one (case (ii) above).
We show in Fig. \ref{fig:m2m4oCRo12} the time development of the real space 
particle distribution of 3 initially circular rings
of stars taken from the initial annulus distribution in Fig. \ref{fig:both_2_o10},
panel with $\Omega_{s,b}=1.2$. The primary spiral wave angular velocity is 
$\Omega_b=1.0$ (as in Fig. \ref{fig:m2m4oCRo07}) placing stars at the CR, but the 
secondary pattern speed is now $\Omega_b=1.2$, which places stars just inside 
the 4:1 OLR. We see in Fig. \ref{fig:both_2_o10}, panel with $\Omega_{s,b}=1.2$,
that the extra velocity dispersion coming from the correlated spirals is about
3 times that due to the individual ones, i.e., $\Delta \sigma_R \approx 3$ (see Eq. 
\ref{eq:delta_sigma}) whereas in the $\Omega_{s,b}=0.7$ panel of the same
figure $\Delta \sigma_R \approx 2$. We consequently expect stronger effect on the
radial diffusion in the case of a secondary spiral wave moving at 
$\Omega_{s,b}=1.2$. This is indeed the case as Fig. \ref{fig:m2m4oCRo12} shows.
\newline

The effect of the addition of a second pattern at a different pattern speed than
the first one is the observed radial diffusion in Figures \ref{fig:m2m4oCRo07} and
\ref{fig:m2m4oCRo12}. For this to work the 2 spirals must be correlated so that 
particles encounter each spiral wave in a systematic manner.  
The radial diffusion of stars is related to the increase of the radial velocity 
dispersion, $\sigma_R$, with time, seen in Figures 
\ref{fig:both_1_o10}-\ref{fig:both_3_o17}. This is most prominent
around the CR or the OLR of the 2-armed spiral wave and the OLR and ILR
of the secondary, 4-armed one.
\newline

\section{Summary and Discussion}
\label{sec:sum}

In this paper we have considered the effect of the simultaneous propagation
of 2 spiral density waves moving at different pattern speeds, on the 
stellar radial velocity dispersion, $\sigma_R$ of a galactic disc. 
The numerical experiments performed concentrated on a primary 2-armed spiral 
wave and a weaker 4-armed one and their effect on the orbits of
particles initially placed into circular orbits. 
This choice of spiral structure is motivated by recent COBE observations
\citep{drimmel}. 

Following the growth of the spiral density waves, we find that 
the variance of the radial velocity resulting from 2 spiral waves 
exceeds that expected from the sum of the variances from each pattern.  
In other words, the velocity dispersion in an integration with both spiral 
density waves is larger than the sum of the effects of the individual spiral 
density waves. The increase in $\sigma_R$ is particularly prominent 
at and near a corotation resonance with a primary, 2-armed spiral
wave, in a combination with a secondary, 4-armed one at and near the OLR and ILR.
Furthermore, we find that $\sigma_R$ continues to increase
near these resonances even after the spiral density waves have ceased to grow.
This suggests that 
the coupling between the 2 spiral waves causes particles to diffuse in real 
space, even though the spiral density wave amplitudes are held fixed. 

\begin{figure*}
\epsscale{0.3}
\plotone{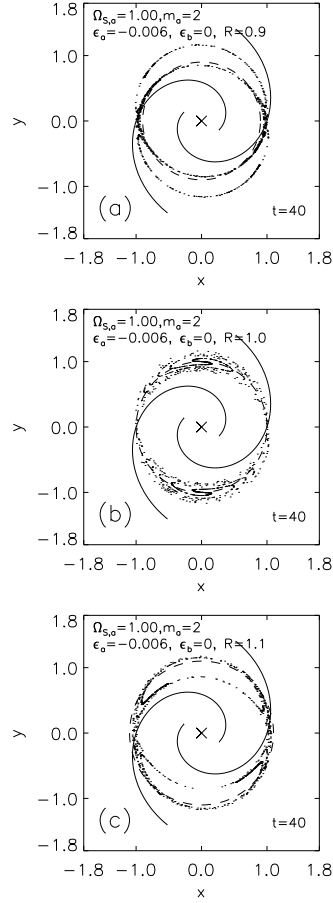}
\figcaption{
A snapshot of the evolution of a ring of stars subjected to a 2-armed spiral 
perturbation grown in the initial 4 rotation periods. The time is $t=40$ rotations. 
The pattern speed is the same for each panel, $\Omega_{s,a}=1.0$, but the radius of 
the initial real-space particle distribution varies as:
(a) $R=0.9$, which places stars just inside the CR; 
(b) $R=1.0$, placing stars exactly at the CR; and
(c) $R=1.1$, placing stars just outside the CR. 
The initial ring in each panel is indicated by the dashed line.
The parameter $\alpha_a$ has the default value of $-6$ which corresponds to a 
pitch angle $p_a\simeq-18\degr$. The curves on which stars lie formed at time 
$t\approx 4$ rotation periods and remained fixed in the reference frame of the 
spiral structure for the rest of the run.
\label{fig:m2_CR}
}
\end{figure*}

\begin{figure*}
\epsscale{0.3}
\plotone{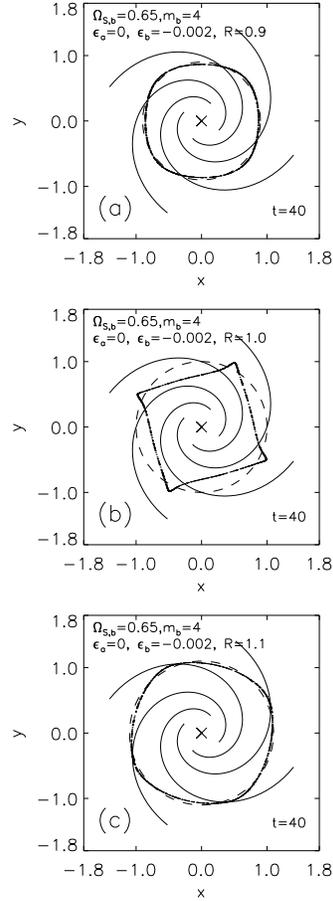}
\figcaption{
A snapshot of the evolution of a ring of stars subjected to a 4-armed spiral
wave at and near the 4:1 ILR. The time is $t=40$ rotations as in Fig. 
\ref{fig:m2_CR}.
The pattern speed is the same for each panel, $\Omega_{s,b}=0.65$, but the radius 
of the initial real-space particle distribution varies as:
(a) $R=0.9$, which places stars just inside the 4:1 ILR;
(b) $R=1.0$, placing stars exactly at the 4:1 ILR; and
(c) $R=1.1$, placing stars just outside the 4:1 ILR.
The initial ring in each panel is indicated by the dashed line.
The parameter $\alpha_b=-12$, corresponding to a pitch angle $p_b\simeq-18\degr$. 
The curves on which stars lie for $R=0.9,1.1$ were approximately
constant in shape and orientation with time, as viewed in the rotation frame with 
the spiral wave. In contrast, exactly at the 4:1 ILR the rectangular curve spins 
clockwise with a period of about 9 rotations.
\label{fig:m4_ILR}
}
\end{figure*}

\begin{figure*}
\epsscale{0.3}
\plotone{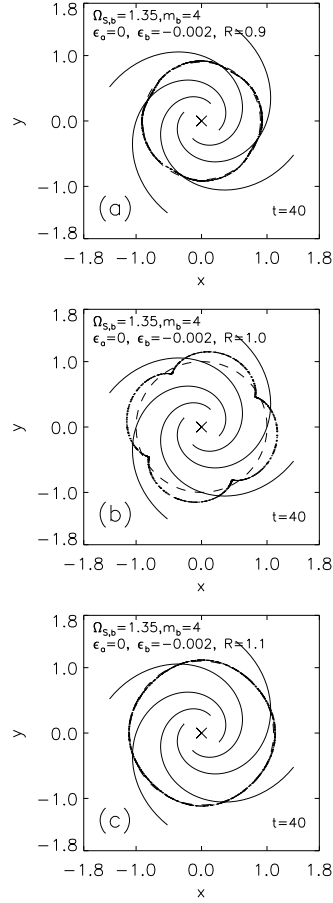}
\figcaption{
Same as Fig. \ref{fig:m4_ILR} but with the angular velocity of the secondary
pattern $\Omega_{s,b}=1.35$, which places stars at the 4:1 OLR at $R=R_0$. 
The pattern speed is the same for each panel, $\Omega_{s,a}=0.65$. 
As in Fig. \ref{fig:m4_ILR}, only when the initial particle distribution placed 
stars exactly at the resonance, panel (b), did the curve on which stars
lie change shape and orientation with time. 
\label{fig:m4_OLR}
}
\end{figure*}

\begin{figure*}
\epsscale{0.72}
\plotone{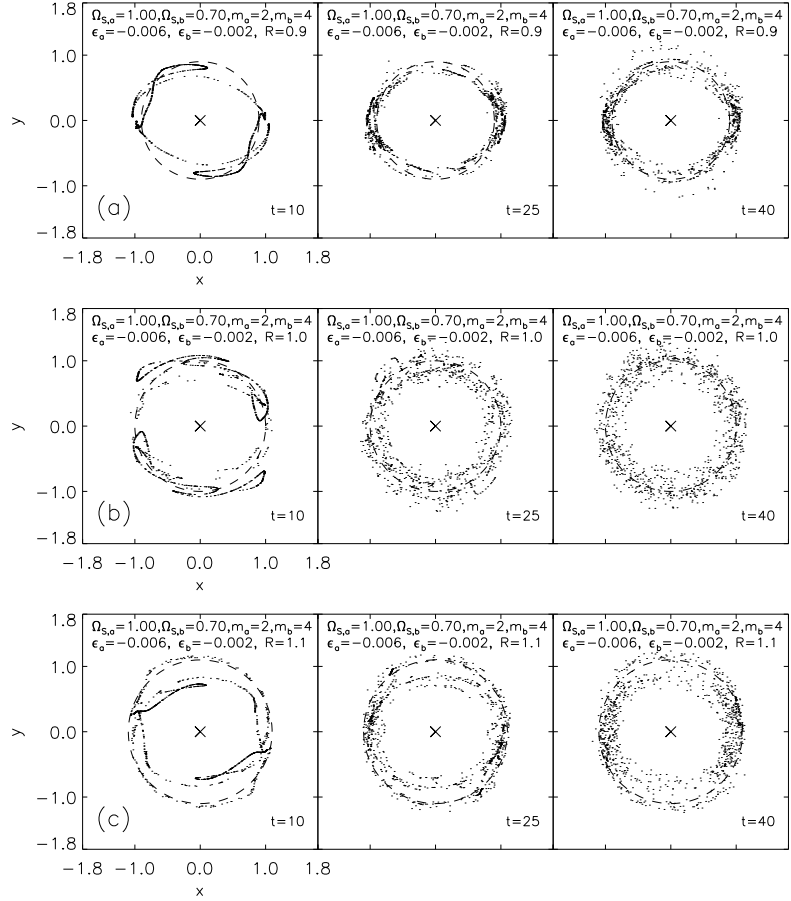}
\figcaption{
Snapshots of the 3-step time evolution of 3 rings of stars subjected to both 
a 2-armed spiral wave near the CR and a 4-armed one near the 4:1 ILR.
This corresponds to the panel with $\Omega_{s,b}=0.7$ in Fig. 
\ref{fig:both_2_o10}. Time develops in each row from left to right.
Note the very different morphology compared to Figures 
\ref{fig:m2_CR}-\ref{fig:m4_OLR}. Stars are observed to diffuse radially.
All parameters of the primary spiral wave in each row are the same as those used
in each row in Fig. \ref{fig:m2_CR}. The difference here is that a second spiral
wave is propagating in addition to the primary one and that we plot
3 steps of the time evolution since changes are taking place unlike in the 
time evolution of the systems shown in Fig. \ref{fig:m2_CR}. 
Rows from top to bottom show particles subjected to 2 spiral wave
perturbations moving at pattern
speeds of $\Omega_{s,a}=1.0$ and $\Omega_{s,b}=0.7$ and initially positioned in
a ring of radius:
(a) $R=0.9$,
(b) $R=1.0$, and
(c) $R=1.1$.
Snapshots are at times $t=10,25,40$. 
The dashed circles show where stars were positioned initially with a 
uniform azimuthal distribution. The crosses show the position of the galactic 
center. 
\label{fig:m2m4oCRo07}
}
\end{figure*}

\begin{figure*}
\epsscale{0.72}
\plotone{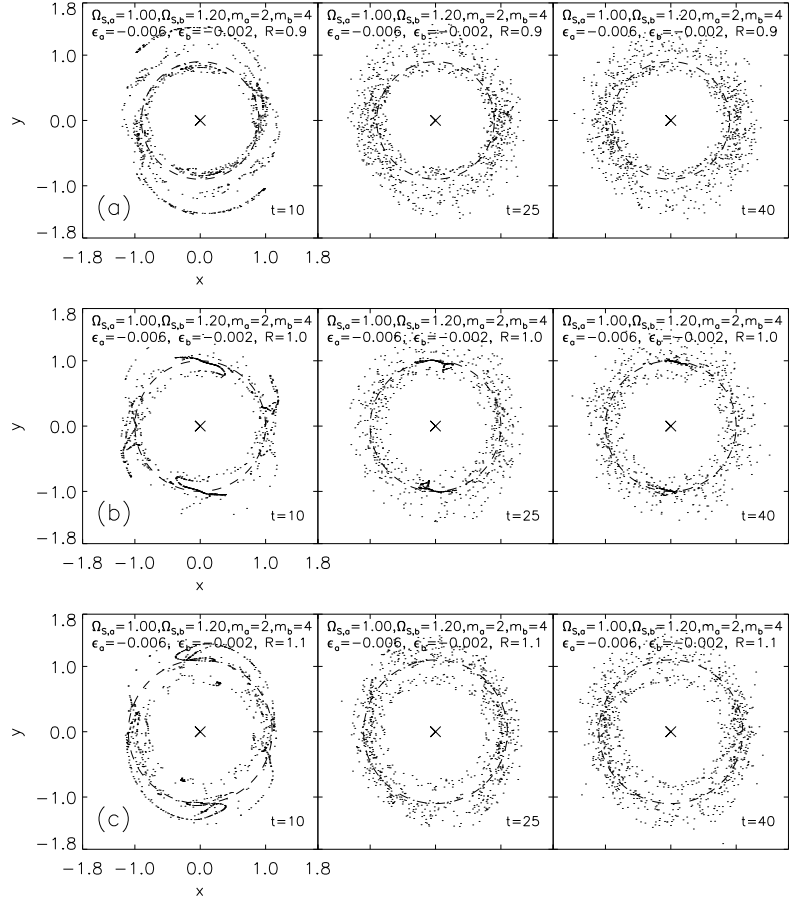}
\figcaption{
Same as Fig. \ref{fig:m2m4oCRo07} but with the secondary pattern speed near the
4:1 OLR, $\Omega_{s,b}=1.2$. This corresponds to the panel with $\Omega_{s,b}=1.2$
in Fig. \ref{fig:both_2_o10}. The radial diffusion of stars is notably increased 
compared to Fig. \ref{fig:m2m4oCRo07}.
\label{fig:m2m4oCRo12}
}
\end{figure*}

We have examined the distribution of stars perturbed by one and two spiral 
density waves. When only one spiral density wave is present, little diffusion in 
phase space occurs. A comparison between these distributions 
(see Figures \ref{fig:m2_CR}-\ref{fig:m2m4oCRo12}) 
shows that diffusion is observed when multiple waves are present.
This comparison confirms our hypothesis that the combined effect of
the two spiral density waves causes stochastic diffusion of stars in phase
space.

Previously studied heating mechanisms have concentrated on heating
from transient spiral density waves (e.g., \citealt{sell_carl,carl_sell,desimone,jenkins,sell_binney}).
However in that case 
changes in the stellar velocity dispersion only occur during the growth or decay
of spiral density waves.  Here we observe an increase in the velocity dispersion
even when the spiral density waves have ceased to grow.  
Hence, this is a different heating mechanism compared to
those explored by these previous investigations.
The heating we see here is similar to that observed at resonances 
in barred galaxies. For example,  \citet{combes} observed stochastic
heating in 3D simulations of barred peanut galaxies. 
We suggest that stochastic heating could also occur in disks of galaxies if
more than one spiral density wave is present.
In this case we could consider the source of the stochastic behavior
the overlap of the two resonant perturbations (e.g., \citealt{murray,lecar}).

We have established the possible combinations of pattern speeds for which
heating is significant. This heating mechanism is strongly 
dependent on the relative velocity between stars and the spiral pattern 
and thus on the galactocentric distance, $R$ (e.g., see Figures 3,4).   
Consequently we expect
strong variations in the velocity dispersion as a function of 
position in the Galaxy.

Future work could consider the role of Giant Molecular Clouds (GMC) in addition 
to the currently explored mechanism by carrying out simulations in 3 dimensions. 
Here we have considered spiral density waves with fixed amplitudes and pitch
angles.  However future explorations could consider spiral structure consistent
with Swing amplification (e.g., \citealt{fuchs}).
Here we have focused on stars initially on circular orbits, however future
explorations could consider stars born in spiral arms. 
Whereas in this paper we used a test-particle approach as was done by 
\citet{desimone}, future explorations can aim to produce self-consistent 
models where the orbits of the stars are consistent with the spiral density 
waves themselves.

\acknowledgements
We wish to thank Gunnar Paesold for useful discussions and for his close 
assistance in software development. We also thank E. Blackman and A. Hubbard
for helpful comments. Support for this work was in part
provided by National Science Foundation grant ASST-0406823,
and the National Aeronautics and Space Administration
under Grant No.~NNG04GM12G issued through the Origins of Solar Systems Program.

{}

\clearpage

\clearpage

\clearpage

\clearpage

\begin{deluxetable}{lcccccccc}
\tablewidth{0pt}
\tablecaption{Parameters describing simulations\label{table:all}}
\tablehead{
\colhead{Figures}        &
\colhead{$\epsilon_a$}   &
\colhead{$\epsilon_b$}   &
\colhead{$\alpha_a$}     &
\colhead{$\alpha_b$}     &
\colhead{$m_a$}         &
\colhead{$m_b$}         &
\colhead{$\Omega_{s,a}$} &
\colhead{$\Omega_{s,b}$}
}
\startdata
\ref{fig:just_e1}
        & -0.006  &        & -6 &      &2& &0.1,0.2,...,1.9,2.0 &                     \\
\ref{fig:just_e2}
     &&-0.001,-0.002,-0.003&    &  -12 & &4&                    &0.5,0.6,...,1.5,1.6  \\
\ref{fig:eps12_heat},\ref{fig:eps12_heat_only}
&-0.006&-0.001,-0.002,-0.003& -6&  -12 &2&4&0.1,0.2,...,1.9,2.0 &0.5,0.6,...,1.5,1.6  \\
\ref{fig:both_1_o10}
        & -0.006  & -0.001 & -6 &  -12 &2&4&  1.0               &0.5,0.6,...,1.5,1.6  \\
\ref{fig:both_2_o10}
        & -0.006  & -0.002 & -6 &  -12 &2&4&  1.0               &0.5,0.6,...,1.5,1.6  \\
\ref{fig:both_3_o10}
        & -0.006  & -0.003 & -6 &  -12 &2&4&  1.0               &0.5,0.6,...,1.5,1.6  \\
\ref{fig:both_3_o17}
        & -0.006  & -0.003 & -6 &  -12 &2&4&  1.7               &0.5,0.6,...,1.5,1.6  \\
\ref{fig:m2_CR}
        & -0.006  &        & -6 &      &2& &  1.0               &                     \\
\ref{fig:m4_ILR}
        &         & -0.002 &    &  -12 & &4&                    &       0.65          \\
\ref{fig:m4_OLR}
        &         & -0.002 &    &  -12 & &4&                    &       1.35          \\
\ref{fig:m2m4oCRo07}
        & -0.006  & -0.002 & -6 &  -12 &2&4&  1.0               &       0.7           \\
\ref{fig:m2m4oCRo12}
        & -0.006  & -0.002 & -6 &  -12 &2&4&  1.0               &       1.2           \\
\enddata
\tablecomments{
Spiral pattern parameters corresponding to simulations
shown in the Figures.
The perturbation strengths $\epsilon_{1}$ and $\epsilon_{1}$ are given in units of 
$V_0^2$, the velocity of a star in a circular orbit at $R_0$. The pattern speeds, 
$\Omega_{s,a}$ and $\Omega_{s,b}$, are in units of $\Omega_0 = V_0/R_0$.
The parameters $\alpha_a$ and $\alpha_b$ set the pitch angles of the spiral arms 
as $m_a \cot(p_a) = \alpha_a$ and identically for $m_b, p_b, \alpha_b$, where 
$p_a$ and $p_b$ are the pitch angles of the 2-armed and 4-armed spirals, 
respectively. Two different initial spatial distributions were used:
(1) a uniform distribution in both $R$ and $\phi$ in the annulus 
($R_0-\Delta R, R_0+\Delta R$) with $\Delta_R=0.3$ (Figures 
\ref{fig:just_e1}-\ref{fig:both_3_o17}); (2) a uniform distribution in the 
azimuthal direction and a delta peak at $R=R_0$, i.e., a ring of stars at $R=R_0$ 
(Figures \ref{fig:m2_CR}-\ref{fig:m2m4oCRo07})
}
\end{deluxetable}


\end{document}